\documentclass{article}
\usepackage[italian]{babel}
\usepackage[svgnames]{xcolor}
\usepackage[globalstyle=1]{mdframed}
\usepackage{amssymb}\usepackage{bbm}
\usepackage{eso-pic}
\usepackage{wrapfig}
\DeclareMathAlphabet{\mathpzc}{OT1}{pag}{m}{it}

\newcommand{\be}{\begin{equation}} 
\newcommand{\ee}{\end{equation}}
\newcommand{\bea}{\begin{eqnarray}} 
\newcommand{\eea}{\end{eqnarray}}
\newcommand{\bc}{\begin{center}}
\newcommand{\ec}{\end{center}}

\newcommand{\à}{\`a}
\newcommand{\è}{\`e}
\newcommand{\é}{\'e}
\newcommand{\ì}{\`\i}
\newcommand{\ò}{\`o}
\newcommand{\ù}{\`u}

\newcommand\AlCentroPagina [1]{\AddToShipoutPicture*{\AtPageCenter{\makebox(0,0){\includegraphics[width =0.9\paperwidth ]{#1}}}}}

\begin{document}
\begin{figure}[!b]
\centering
\includegraphics[width=2cm]{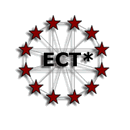}
\end{figure}
\AlCentroPagina{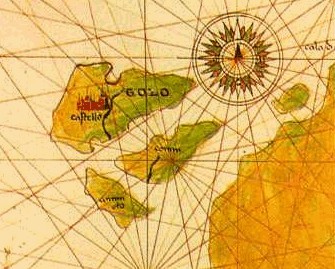}
\title{ \LARGE ESPLORAZIONE DELLA TERRA INCOGNITA DELLA FISICA NUCLEARE:
nuclei con alone, nuclei "a grappoli", nuclei borromeani e altre stranezze esotiche! } 
\author{ Lorenzo Fortunato -ECT* \\
European Center for Theoretical Studies\\
 in Nuclear Physics and Related Areas \\
Strada delle Tabarelle 286,  I-38123 Villazzano (TN), Italy \\
fortunat@pd.infn.it }  
\maketitle

\begin{abstract}
{ \tt \Large Se lo stato attuale delle conoscenze teoriche e sperimentali della Fisica Nucleare si potesse ridurre a una carta geografica o un mappamondo, allora la nostra epoca si potrebbe paragonare per certi versi alla fine del Medioevo e per altri al Rinascimento: epoche in cui, accanto alla conoscenza consolidata del Vecchio mondo, ci furono grandi e rivoluzionarie spedizioni come la scoperta dell'America o il periplo del Capo di Buona Speranza, esplorazioni geografiche volte al raggiungimento dei limiti che nel corso dei secoli hanno portato a scoperte scientifiche sempre pi\ù avvincenti in terre sempre pi\ù lontane, culminate solo nel ventesimo secolo con la completa mappatura di tutte le terre emerse. La Fisica Nucleare sta vivendo un'avventura analoga sul piano della contrapposizione tra la conoscenza acquisita sui nuclei stabili e la scoperta delle sorprendenti propriet\à esotiche dei nuclei instabili, pi\ù difficili da produrre, da studiare e da comprendere.}
\end{abstract}

\section{ INTRODUZIONE}
Come avviso ai naviganti posso dire che con il termine Fisica Nucleare non intendo parlare delle applicazioni tecnologiche che godono di una grande eco pubblica (pertanto tutti coloro che speravano di leggere di energia nucleare e bombe atomiche rimarranno delusi),  bens\ì di quella branca fondamentale delle scienze fisiche che si occupa della descrizione e comprensione delle leggi che governano i fenomeni naturali che avvengono tra le particelle che costituiscono il nucleo atomico, a distanze dell'ordine di pochi milionesimi di miliardesimi di metro (ovvero di alcuni fermi o femtometri, 1 fm = 10$^{-15}$m). E' questa una precisazione doverosa, perch\é in quasi nessun campo del sapere umano come in fisica nucleare, la stragrande maggioranza del pubblico di non specialisti (e tra questi in modo evidente i nostri politici, che snobbano la ricerca fondamentale in favore delle applicazioni ``ripagabili'' in tempi brevi) ignora la differenza tra applicazioni e scienza fondamentale, che rimangono a mio avviso due capitoli tanto ben distinti quanto inestricabilmente interconnessi \footnote{Tranne in casi di evidente serendipit\à, i passi in avanti tecnologici sono il frutto di studi sperimentali e teorici effettuati in ambito accademico, magari decenni o secoli prima. Si pensi alla radio, sarebbe stata possibile senza il corposo trattato sull'elettromagnetismo scritto da Maxwell nel 1863? Pertanto sarebbe una follia affossare lo studio tout court delle discipline pure in favore di quelle applicate: la nostra societ\à ne pagherebbe le conseguenze sul lungo periodo!}. Un'altra precisazione \è d'obbligo: spesso si pensa in maniera superficiale che tutto ci\ò che \è ``nucleare'' (e lo stesso potrebbe valere per altri aggettivi come ``atomico'', ``gravitazionale'', ``elettromagnetico'', ``particellare'' e via dicendo) sia qualcosa di artificiale, qualcosa che pu\ò far del male, qualcosa che viene prodotto e studiato in laboratorio da personale scientifico in camice bianco. Questa visione na\"\i f di ispirazione hollywoodiana \è molto lontana dal vero: la fisica \è lo studio della natura, del mondo per come lo percepiamo ed analizziamo con la ragione, trascrivendone le leggi con un linguaggio matematico, e la fisica nucleare \è semplicemente quella sottodisciplina che si occupa del nucleo atomico, un oggetto che si trova nella natura, tanto quanto i fiori, le salamandre ed i cristalli di quarzo, sebbene ad una scala talmente piccola da sfuggire all'osservazione diretta per mezzo dei nostri sensi.

Intendo fornire, con il presente articolo, una panoramica pedagogica sugli sviluppi pi\ù recenti e sorprendenti di questa branca delle scienze.

\section{ IL CENTRO DELL'ATOMO}
\begin{figure}[!t]
\centering
\includegraphics[width=8cm]{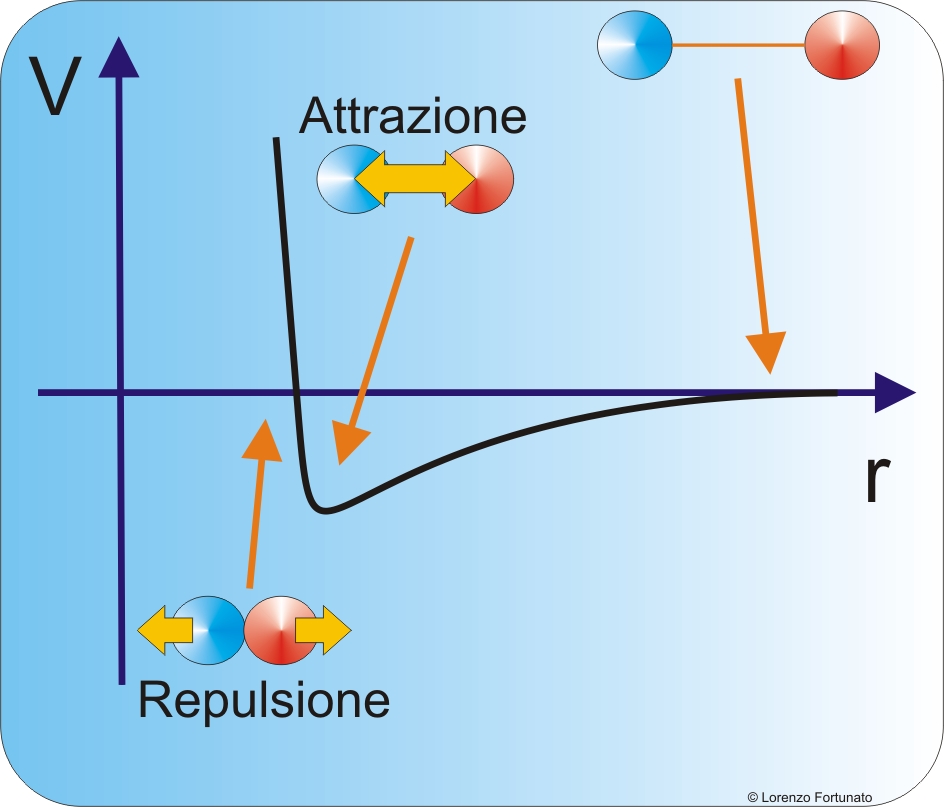}
\caption{Potenziale nucleone-nucleone schematico in funzione della distanza: per grandi distanze non c'\è praticamente interazione tra i nucleoni, alle distanze intermedie c'\è attrazione (potenziale negativo), che tiene i nucleoni ad una certa distanza e di conseguenza tiene uniti anche i nuclei atomici, mentre alle corte distanze c'\è una forte repulsione (potenziale positivo) che impedisce ai nucleoni di collassare gli uni sugli altri. Le lunghezze tipiche in gioco sono dell'ordine del femtometro, anche detto fermi ed abbreviato in fm, che corrisponde al milionesimo di miliardesimo di metro! (Rielaborazione di una figura tratta da \cite{heyde}).}
\label{pot}
\end{figure}

La materia ordinaria (ebbene s\ì, anche le salamandre!) \è costituita da atomi e molecole, formate dall'unione di due o pi\ù atomi. Le leggi che governano i processi che sperimentiamo attorno a noi e dentro di noi ogni giorno sono soprattutto leggi chimiche, relative alle proporzioni degli atomi e alle reazioni che cambiano queste proporzioni. Esse avvengono alla scala di lunghezze del decimo di miliardesimo di metro e alla scala di energie dell'elettronvolt, un'unit\à di misura piccolissima per noi umani, che corrisponderebbe all'energia necessaria a sollevare di un miliardesimo di centimetro un piccolo peso da circa 1,6 miliardesimi di grammo. 
Da circa un secolo ovvero dalla scoperta del nucleo atomico ad opera di Ernest Rutherford e colleghi si sa che l'atomo non \è indivisibile, ma \è formato da un nocciolo di carica elettrica positiva, duro, denso e contenente quasi tutta la massa dell'atomo, e da una nuvola, leggera e poco densa, di elettroni con carica elettrica negativa che gli orbita attorno. La scoperta del nucleo ha permesso una riclassificazione degli elementi chimici sulla base della carica elettrica del nucleo (il numero atomico, indicato con Z) che corrisponde al numero di protoni presenti nel nucleo. I protoni sono i costituenti basilari \footnote{Ai regimi di energia tipici della struttura nucleare ($<$20-30MeV) la composizione dei nucleoni in termini di quark non viene presa in esame: i nucleoni possono essere considerati a tutti gli effetti come costituenti fondamentali del nucleo. Diverso \è il caso in cui si stiano considerando energie pi\ù elevate o si superi la soglia di massa del pione ($ \sim$ 135-140 MeV): in tal caso \è necessario affinare la descrizione dei nucleoni e considerare il fatto he essi sono particelle composte.} del nucleo insieme ai neutroni, scoperti da J.Chadwick nel 1932. Questi ultimi sono neutri, come suggerisce il nome, cio\è non posseggono carica elettrica ed hanno all'incirca la stessa massa dei protoni (in effetti un pochino di pi\ù). Mentre il protone \è una particella stabile (o meglio per quel che si sa non decade in nulla di pi\ù semplice, nemmeno se lo osservassimo per tempi comparabili con l'et\à stimata dell'universo), il neutrone libero, nel vuoto, decade in circa 15 minuti in un protone, un elettrone ed un antineutrino, per via del decadimento beta, dovuto alle cosiddette forze deboli. 
Collettivamente, senza far differenza per quanto riguarda la carica elettrica, protoni e neutroni vengono detti nucleoni, veri e proprio mattoncini Lego del nucleo, il cui numero viene detto peso atomico (indicato con A).
All'interno dei nuclei atomici i nucleoni sono soggetti ad una forza a corto raggio (la cosiddetta forza forte) che \è praticamente nulla a distanze molto grandi e diventa molto attrattiva a distanze intermedie dell'ordine di alcuni fermi (si veda la figura \ref{pot}). A distanze molto brevi invece, questa forza diventa fortemente repulsiva e ci\ò spiega come mai i nucleoni nel nucleo tendono ad aggregarsi, ma non a collassare gli uni sugli altri. 

Inoltre, per i soli protoni, \è in opera anche la forza di repulsione elettrostatica di Coulomb che prevede che cariche dello stesso segno si respingano in maniera inversamente proporzionale alla loro distanza. Se non fosse per la presenza "collante" dei neutroni, i protoni da soli non riuscirebbero ad attrarsi in maniera stabile. Pertanto, mentre il nucleo dell'atomo di idrogeno \è formato semplicemente da un solo protone, l'atomo di elio-4 ha un nucleo che consiste di due protoni, ma anche di due neutroni, necessari alla stabilit\à del sistema. All'altro estremo della tavola degli elementi il piombo-208, uno dei pi\ù massicci nuclei stabili, consta di 82 protoni e ben 126 neutroni!

\section{ ISOTOPI E TAVOLA DEI NUCLIDI}
Va precisato che mentre la chimica {\it fin de si\ècle} e le teorie atomiche di oltre un secolo or sono, tentavano di classificare i diversi elementi basandosi solo sulle propriet\à fisico-chimiche, come densit\à,  punto di fusione o comportamento nelle reazioni chimiche, una prima rivoluzione si ebbe associando la posizione nella tavola alla carica elettrica del nucleo, che viene detta numero atomico Z. In seguito le scoperte della fisica nucleare hanno permesso di precisare il ruolo specifico dei vari costituenti e le loro proporzioni, pertanto si \è visto che di ogni elemento chimico (quindi per un fissato numero di protoni) esistono diverse varianti, chiamate isotopi, che presentano (quasi) le medesime propriet\à chimiche, ma il cui nucleo pu\ò contenere un numero di neutroni variabile: questa variazione del numero di neutroni determina propriet\à fisiche diverse di tipo strutturale, come l'energia di legame, cio\è l'energia necessaria ad estrarre una particella dal nucleo,  e anche di comportamento nelle reazioni nucleari, come ad esempio una maggiore o minore tendenza alla fusione con un altro nucleo.
Partendo dal nucleo pi\ù semplice, l'idrogeno (Z=1), sappiamo, come detto, che esso \è caratterizzato da un solo protone, ma possiede alcuni isotopi, come il deuterio, formato da un protone ed un neutrone, o il tritio (si legge ``trizio'' poich\é si rif\à alla pronuncia latina), mentre l'elio (Z=2), caratterizzato da due protoni, si pu\ò incontrare in natura sotto forma di elio-3, contenente due protoni e un neutrone, o sotto forma di elio-4, con due protoni e due neutroni.  Quando si parla di isotopi si deve specificare che essi formano stati legati, ma non necessariamente stabili, cio\è immutabili nel tempo. Un certo insieme di particelle forma uno stato legato quando \è necessario dover fornire energia per poter estrarre una delle particelle dal mucchio.  Viceversa lo stato non \è legato se una particella (o un insieme di particelle) si allontana spontaneamente dal resto. Tuttavia pu\ò darsi il caso che uno stato legato sia instabile, cio\è che, dopo un certo tempo caratteristico del sistema, abbia la tendenza a decadere in un altro stato legato dello stesso sistema che si trovi ad un'energia pi\ù bassa o addirittura a trasformarsi in un nuovo sistema, formato da una diversa miscela di neutroni e protoni (\è questo il caso dei decadimenti $\alpha$, $\beta+$ e $\beta-$, si veda inserto sui decadimenti).
Gli elementi chimici conosciuti al giorno d'oggi sono poco pi\ù di un centinaio: se li ordiniamo per numero atomico, 91 dei primi 92 si trovano comunemente in natura nascosti nei minerali che affiorano sulla crosta terrestre con l'eccezione del tecnezio (Z=43), che \è stato prodotto in laboratorio a partire dal molibdeno. Va poi ricordato che solo 90 di questi 91 possiedono almeno un isotopo stabile: infatti, il promezio (Z=61) non ha isotopi stabili ed esiste in natura solo in tracce minime\footnote{Certe stime, ricavate da esperimenti ultraprecisi di decadimento dell'uranio nel minerale pechblenda, indicano che in tutta la crosta terrestre probabilmente non c'\è pi\ù di mezzo chilogrammo di promezio. Questo ovviamente non tiene in conto che nuovi giacimenti possano venire scoperti in futuro (per esempio sotto i ghiacci!), ma d\à un'indicazione della rarit\à di questo elemento.}. Il novantaduesimo \è l'uranio e al momento della scrittura di questo articolo un'altra ventina di elementi transuranici superpesanti sono stati prodotti con varie tecniche negli acceleratori di particelle, fino ad arrivare al darmstadio, al roentgenio e al copernicio, che hanno numeri atomici 110, 111 e 112. A quest'ultimo il nome \è stato assegnato dalla IUPAC, ovvero International Union of Pure and Applied Chemistry, che \è l'organizzazione preposta a questo tipo di attivit\à, solo di recente (Feb. 2010). 
Il numero totale di isotopi conosciuti, appartenenti a questi 112 elementi, \è superiore a 4000, di cui circa un decimo \è piuttosto comune in natura, mentre il resto \è raro o del tutto assente nel nostro ambiente. Lo si pu\ò riscontrare solo in ambienti estremi come le stelle o ottenere in quantit\à minuscole come prodotto di reazioni nucleari eseguite in laboratorio.  

\begin{figure}[!t]
\centering
\includegraphics[clip=,width=12cm]{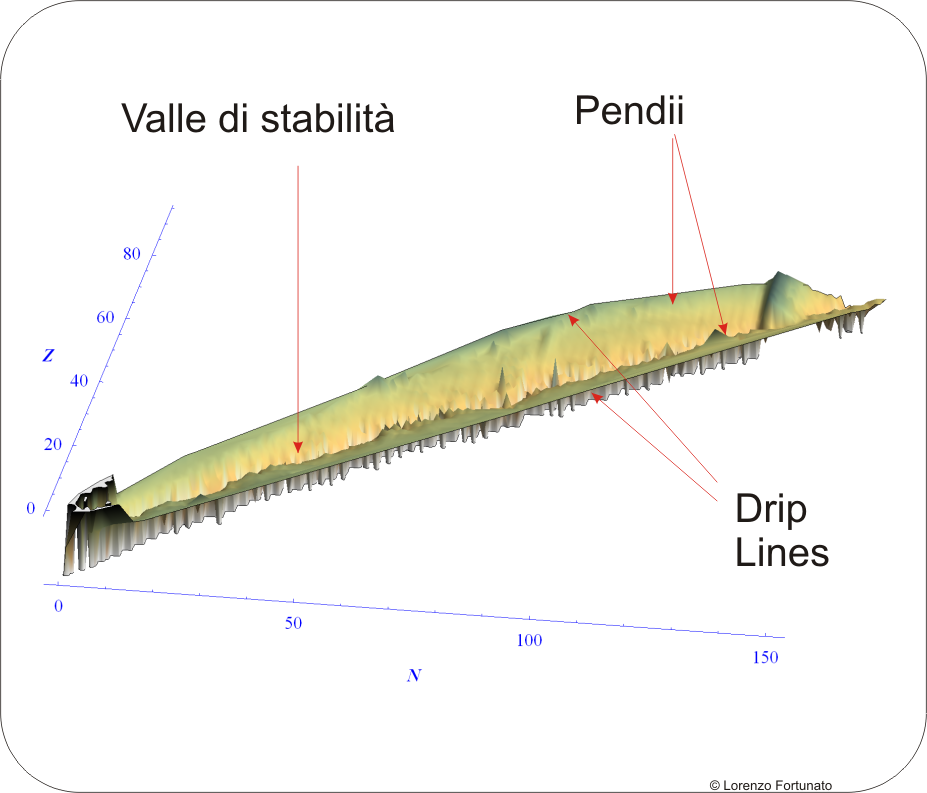}
\caption{Una rappresentazione ``topografica'' tridimensionale della valle di stabilit\à. In funzione del numero di protoni (Z) e del numero di neutroni (N) la tendenza a decadere \è tanto maggiore quanto pi\ù si sale di quota. La valle in questo schema poggia sul piano. I nuclei sono tanto pi\ù instabili quanto pi\ù ci si sale lungo i pendii e ci si avvicina alle ``drip lines'', ovvero alle linee oltre la quali i singoli nucleoni sono slegati.}
\label{valle}
\end{figure}
Tutti questi isotopi si possono rappresentare in una tavola, detta tavola degli isotopi, che \è una vera e propria cartina geografica del mondo del nucleo atomico. Gli assi orizzontali e verticali rappresentano rispettivamente il numero di neutroni e di protoni, e gli isotopi sono piazzati nelle caselle corrispondenti. I nuclei stabili occupano la porzione centrale della tavola e almeno per un po' seguono approssimativamente la diagonale: ci\ò \è dovuto al fatto che, finch\é il numero di protoni \è piccolo, la configurazione in cui ci sono tanti neutroni quanti protoni \è quella con l'energia pi\ù bassa e quindi la pi\ù stabile.  Man mano che il numero di protoni aumenta, aumenta anche la forza di repulsione elettromagnetica e quindi \è necessaria una maggior quantit\à di neutroni per fare da collante. Questa zona \è detta valle di stabilit\à e, volendo portare all'estremo il paragone con una cartina topografica, si pu\ò pensare che le due aree separate dalla valle siano come delle rive in una vallata, la cui quota rappresenta una tendenza all'instabilit\à. Pi\ù ci si allontana dalla valle di stabilit\à verso destra (sinistra) aggiungendo (o rispettivamente sottraendo) neutroni, pi\ù gli isotopi che si incontrano sono instabili per decadimento $\beta$ e pi\ù la loro vita media diventa piccola. Questo fino ad un punto in cui aggiungendo (o sottraendo) un neutrone non si riesce pi\ù a formare un isotopo che possa rimanere stabile per un tempo qualsivoglia piccolo. Le linee che delimitano queste aree vengono dette linee di caduta o precipitazione (in realt\à tutti usano il termine anglosassone di ``drip lines''), poich\é, se riprendiamo il paragone topografico, \è come se salendo sul pendio della vallata si arrivasse ad un bordo oltre il quale, proprio come credevano i geografi antichi, ci sia un baratro infinitamente profondo! Le zone tra la valle di stabilit\à e le ``drip lines'' sono popolate di nuclei molto ricchi (o rispettivamente poveri) di neutroni che vengono detti nuclei esotici per la loro rarit\à, o addirittura assenza, in natura e per la difficolt\à di produrne fasci intensi come invece avviene per i nuclei stabili. Un modo matematicamente pi\ù preciso di definire le ``drip lines'' di neutrone (protone) \è quello di dire che sono le linee alle quali l'energia di separazione di un neutrone (protone) \è nulla. L'energia di separazione \è l'energia necessaria a portare una particella a distanza infinita dal centro di massa del nucleo.
Proseguendo lungo la vallata, che come detto si discosta dalla diagonale, si arriva a un punto in cui la vallata stessa termina in una zona in salita, oltre il piombo, infatti, non esistono pi\ù nuclei stabili: gli elementi in questa regione tendono a decadere per decadimento $\alpha$ formando nuclei figlio che ritornano in prossimit\à della valle di stabilit\à. Molti fisici sono convinti, per ragioni la cui spiegazioni richiederebbe risorse che vanno al di l\à dello scopo di questo articolo, che al di sopra di questa zona, vi sia un'isola di stabilit\à \footnote{E' interessante notare che ci sono due versioni speculari di questo paragone geografico: la versione usata nel testo, in cui vi è una vallata contornata da due rive che termina in salita ed una, per cos\ì dire al negativo, in cui c'\è una specie di catena montuosa al centro la cui cresta termina con una discesa. In questa seconda versione le drip lines si possono immaginare come il livello del mare che circonda la nostra catena montuosa e l'isola di stabilit\à \è un'isola vera e propria che sorge dalle acque. Nella prima versione, invece, sarebbe più opportuno immaginare una ``pozza di stabilit\à '' o una depressione! Io credo che il termine isola abbia prevalso, inconsciamente, per dare l'idea di un approdo, di una meta da raggiungere.}, ovvero una regione di massa e carica in cui alcuni isotopi potrebbero avere vite medie significativamente pi\ù alte di quelli che li circondano o perfino essere stabili! La ricerca di questi isotopi, anche se estremamente difficile, rappresenta un importante ed affascinante tema di sviluppo. I fisici nucleari, come moderni Ulisse, cercano di spingersi al di l\à delle colonne d'Ercole dei nuclei superpesanti.

\section{ I NUCLEI ESOTICI}
\subsection*{ALONI, GRAPPOLI, PELLE DI NEUTRONI E MOLECOLE NUCLEARI}
Tornando al nostro argomento principale, si \è visto negli ultimi venti o trent'anni, grazie a sofisticati acceleratori e ad apparati sperimentali sempre pi\ù perfezionati, che si possono formare fasci di nuclei esotici abbastanza intensi da permettere lo studio quantitativo della struttura e delle reazioni di questi bizzarri isotopi. I fisici, come padri ansiosi per la nascita dei figlioletti, si sono sbizzarriti con i nomi pi\ù strani, che spesso hanno a che vedere con le inusuali caratteristiche strutturali o con il comportamento irregolare in certe reazioni nucleari \cite{alka}.

\begin{figure}[!t]
\centering
\includegraphics[clip=,width=5cm]{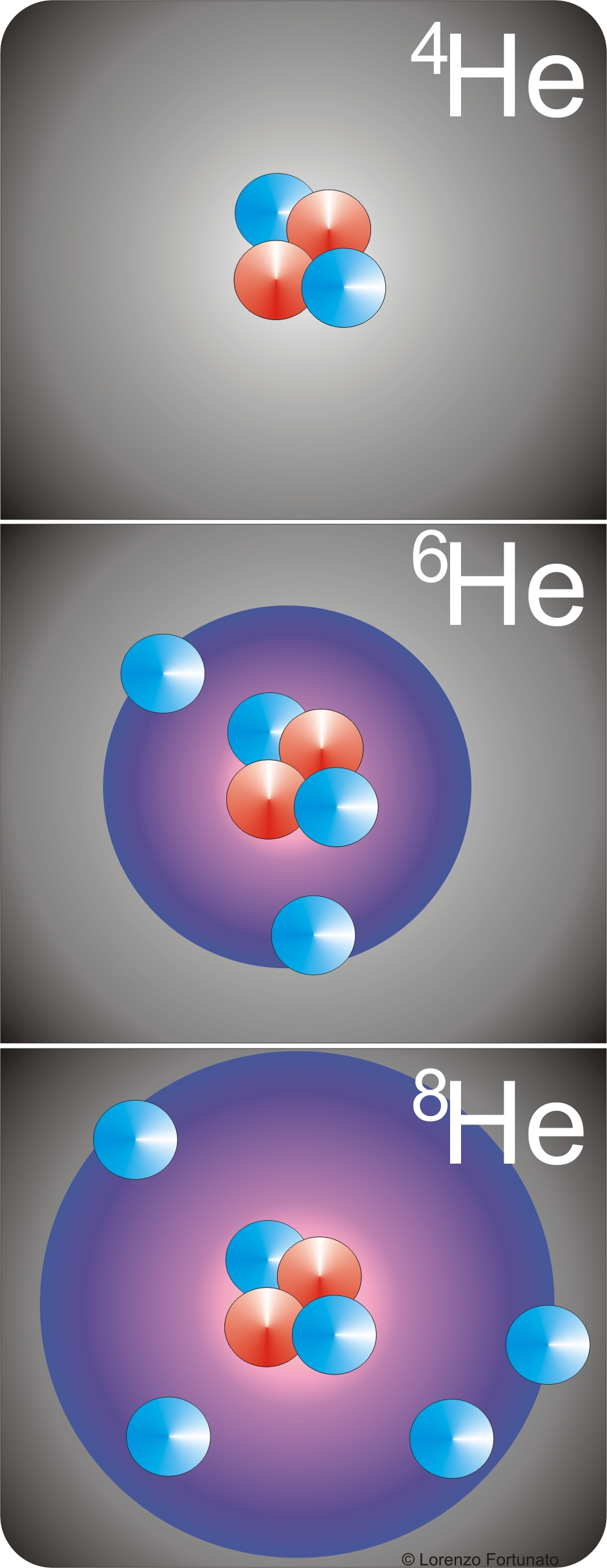}
\caption{Rappresentazione schematica dei nuclei di elio-4, elio-6 ed elio-8. Mentre il primo \è molto fortemente legato, spazialmente compatto e caratterizzato da una zona superficiale netta, questi ultimi due sono solo debolmente legati, possiedono una forte estensione spaziale e una zona superficiale, detta alone, molto diluita e diffusa.}
\label{he}
\end{figure}

Un esempio per tutti di nuclei esotici dalle intriganti propriet\à \è fornito dagli isotopi dell'elio. Come abbiamo gi\à ricordato in natura si incontrano l'elio-3 e l'elio-4 (detto anche particella a) e manca l'elio-2 che corrisponderebbe ad avere solo 2 protoni senza alcun neutrone: questo isotopo non esiste perch\é i protoni si respingono per via della repulsione elettrostatica pi\ù di quanto si attraggano per interazione forte. Poi ci sono isotopi con un pi\ù alto numero di neutroni (si veda la figura \ref{he}): l'elio-5 non \è stabile, ma l'elio-6 s\ì. Allo stesso modo l'elio-7 non \è stabile, ma l'elio-8 s\ì. E' piuttosto complicato arrivare a capire con precisione i dettagli di questo curioso fenomeno, ma, semplificando un pochino e tralasciando le considerazioni che deriverebbero dal modello a shell, si pu\ò dire che, mentre i primi quattro nucleoni (2 protoni e 2 neutroni) occupano una porzione di spazio vicina al centro sovrapponendosi e interagendo intensamente, il terzo neutrone si trova su un'orbita leggermente pi\ù esterna e non interagisce abbastanza fortemente con le altre quattro particelle da formare un insieme stabile. Ma se ammettiamo che due neutroni possano venire a trovarsi su quest'orbita esterna, allora, sebbene ognuno di essi separatamente non interagisca abbastanza con la particella $\alpha$, la loro mutua interazione, detta di pairing o di accoppiamento, fa s\ì che essi formino un legame stabile con gli altri quattro nucleoni. Semplificando ancor di pi\ù, un neutrone avrebbe 4 interazioni con l'elio-4 \footnote{  I chimici  ``alla Lewis'' direbbero legami, ma questa parola \è inappropriata in questo contesto poich\é in realt\à non si formano strutture rigide come le molecole, ma piuttosto abbiamo a che fare con delle gocce di fluido quantistico!} , ma due neutroni ne avrebbero 8 con l'elio-4 pi\ù una tra di loro ed il bilancio totale di 9 interazioni permette la formazione di un legame stabile. Lo stesso fenomeno avviene al livello successivo e ci\ò spiega qualitativamente la formazione di uno stato legato nell'elio-8. Le energie di legame in questi isotopi pesanti dell'elio sono decisamente piccole (su scala nucleare), perci\ò si parla di nuclei debolmente legati. Per capirne l'entit\à si pensi che l'energia necessaria a estrarre un neutrone da una particella $\alpha$ \è maggiore di 20 MeV (MegaelettronVolt, un'unit\à di misura dell'energia adatta alla fisica nucleare che \è un milione di volte pi\ù grande dell'elettronvolt citato in precedenza per le energie tipiche della chimica), il caso tipico di un nucleo medio pesante prevede un'energia di estrazione di 6-7 MeV, mentre nel caso dell'elio-6, \è necessaria solo una frazione di MeV, una quantit\à misera rispetto al tipico bilancio energetico dei nuclei.

\begin{figure}[!t]
\centering
\includegraphics[clip=,width=10cm]{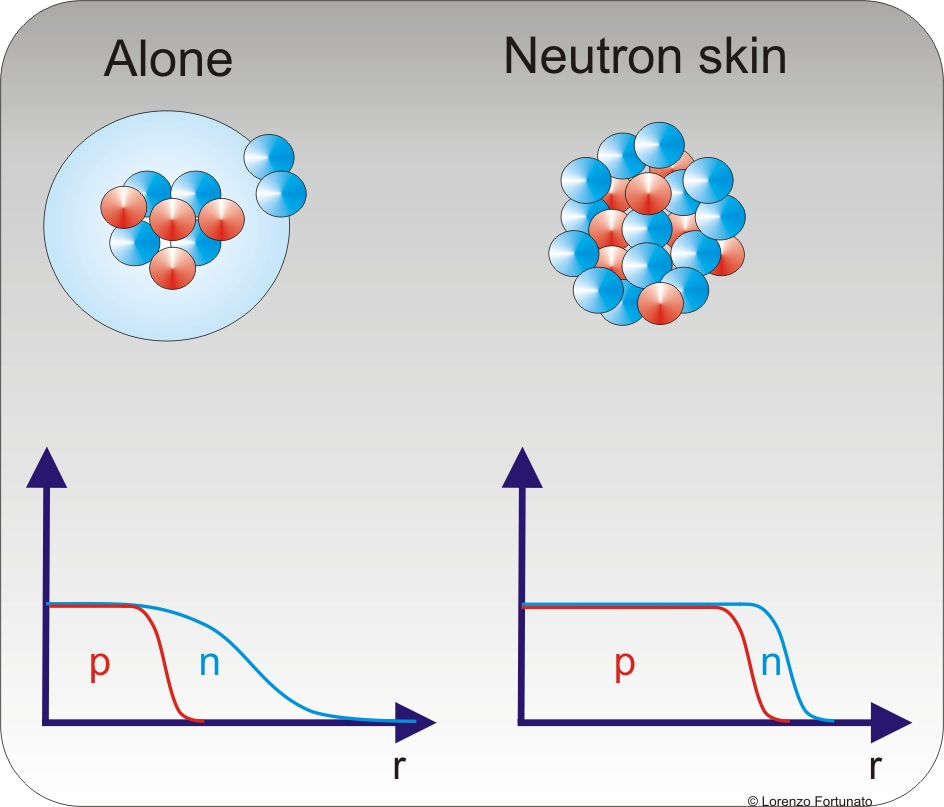}
\caption{Andamento schematico delle densit\à di protoni (in rosso) e neutroni (in blu) in nuclei con alone e in nuclei con pelle di neutroni ("neutron skin"). I due neutroni dell'alone di sinistra occupano un volume molto grande e conseguentemente sono molto diffusi e poco densi, mentre a destra le densit\à di neutroni e protoni hanno la stessa forma, ma i neutroni, essendo pi\ù numerosi prevalgono nella regione esterna con una densit\à paragonabile a quella centrale.}
\label{skin}
\end{figure}

Non solo l'elio-4 \è pi\ù intensamente legato, ma \è anche pi\ù denso e compatto del suo esotico fratellino. Il suo raggio di materia \è di soli 1.7 fm, mentre il raggio dell'elio-6 \è di circa 2.6 fm, ovvero tanto quanto un nucleo senza alone formato da 10-11 nucleoni. La grande zona esterna che ospita solo due neutroni viene detta alone nucleare ed \è caratterizzata dalla bassa densit\à rispetto al nocciolo centrale. 
Un esempio ancor pi\ù sorprendente \è fornito dal nucleo litio-11 che avendo un raggio di 3.5 fm ha, di conseguenza, circa lo stesso volume di un nucleo di magnesio-25, pur possedendo meno della met\à delle particelle di quest'ultimo. Il magnesio per\ò ha una densit\à pressappoco costante dal centro fino alla superficie, mentre il litio-11 \è ``quasi'' vuoto. Anche il concetto stesso di superficie, che ha senso per nuclei ordinari \footnote{Ad essere precisi i nuclei sono oggetti quantistici, i cui costituenti non sono raffigurabili come oggetti del mondo alla scala umana: il concetto di superficie o volume o di estensione spaziale, che \è chiaro per una pallina non ha senso, se non in maniera approssimata per una ``nuvoletta''. Quando si ha a che fare con distribuzioni di probabilit\à o distribuzioni di una qualsivoglia grandezza fisica sarebbe pi\ù corretto parlare di valore di aspettazione di un operatore opportuno sui corrispondenti stati quantici.} in cui la densit\à cala rapidamente a zero in una zona superficiale molto stretta, perde significato con questi nuclei esotici, perch\é non \è pi\ù possibile identificare una zona di cambio repentino, ma piuttosto bisogna immaginarsi una nuvoletta, grigia al centro, che sfuma lentamente verso l'esterno (si veda la parte sinistra della figura \ref{skin}).

Il fatto che isotopi diversi di uno stesso elemento potessero avere raggi che proprio non seguono la semplice legge di Fermi ($R=r_0 A^{1/3}$, dove R \è il raggio nucleare, $r_0$=1.2 fm \è un parametro e A \è la massa) era gi\à stato notato nel 1985 con gli esperimenti di Tanihata dai quali si deduceva il raggio dei nuclei della catena del litio. 
I nuclei con alone di singola particella, proposti teoricamente per la prima volta da Hansen e Jonson nel 1987, sono pi\ù semplici di quelli con due particelle nell'alone, ma anche pi\ù rari: per esempio il berillio-11 ed il carbonio-19 hanno una struttura con un alone di neutrone, mentre il boro-8 o il fluoro-17 hanno un alone di protone.  Nel caso del berillio, l'energia di legame \è di solo mezzo MeV e l'estensione spaziale dell'alone \è attorno ai 5-6 fm, molto maggiore del raggio tipico di altri nuclei nella stessa regione di massa. L'alone implica anche una forte polarizzabilit\à elettrica, ovvero la possibilit\à di indurre una dissociazione per mezzo del campo elettrico di un altro nucleo che venga a passare nelle vicinanze.

\begin{figure}[!t]
\centering
\includegraphics[clip=,width=6cm]{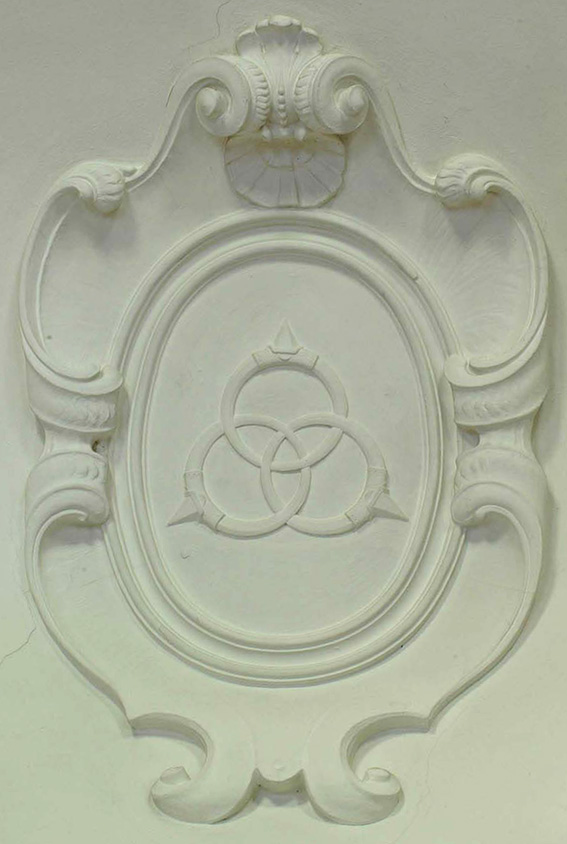}
\caption{Uno degli stemmi araldici della casa dei Borromeo, detto anche nodo borromeano. Secondo gli storici simboleggia l'amicizia tra le potenti famiglie Borromeo, Sforza e Visconti. Secondo altri \è generalmente \è un richiamo religioso alle tre virt\ù teologali o all'unit\à della trinit\à. Nell'ambito scientifico \è stato coniato il termine di nuclei borromeani per riferirsi a quei sistemi legati in cui \è possibile individuare tre sottosistemi, tali che, quando uno dei tre viene rimosso, gli altri due non sono in grado di formare un legame stabile, ma si dissociano. Lo stesso accade sfilando o spezzando uno degli anelli: gli altri due non stanno uniti.}
\label{borr}
\end{figure}

\begin{figure}[!t]
\centering
\includegraphics[clip=,width=5cm]{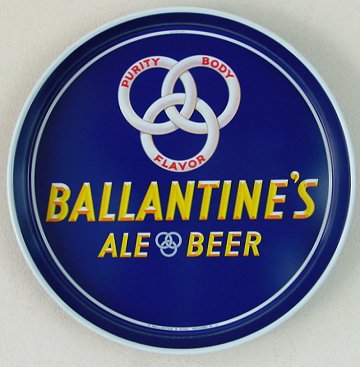}
\caption{Un altro ``nobile'' esempio dell'utilizzo dei tre cerchi borromeani come simbolo per una marca di birra americana, la Ballantine Ale, fin dal 1879. Si narra che il simbolo ed il motto ``Purity, Body, Flavor'' furono ispirati a Peter Ballantine (scozzese emigrato in America nel 1820) dagli anelli di condensa lasciati dalla bottiglia su un tavolo. }
\label{ball}
\end{figure}

A proposito del fatto che l'elio-5 non si riscontra in Natura, ma l'elio-6 s\ì, i fisici hanno coniato una definizione curiosa: si dicono borromeani quei nuclei legati formati da tre sottosistemi, tali che, quando uno dei tre sottosistemi viene rimosso, gli altri due non riescono a formare uno stato legato e si dissociano immediatamente. Il nome inconsueto, reminescente della ricca e potente famiglia milanese dei Borromeo, il cui pi\ù famoso rampollo \è il notissimo S.Carlo Borromeo, deriva dal fatto che nello stemma nobiliare di questa casata sono raffigurati, tra le altre cose, anche tre anelli stilizzati, simbolo dell'amicizia tra le famiglie dei Borromeo, degli Sforza e dei Visconti, incatenati in maniera tale che se uno dei tre viene spezzato gli altri due non possono in alcun modo stare uniti (provate a realizzarlo con dello spago e del nastro adesivo!).  Questi intriganti cerchi (si vedano le figure \ref{borr} e \ref{ball}), che vengono spesso chiamati nodi borromeani o anelli borromeani ritornano in matematica, in fisica, in chimica, in biologia e in moltissimi altri ambiti del sapere umano\cite{ring} come gli origami, la teologia cristiana o la filosofia orientale e risalirebbero addirittura a iscrizioni runiche degli antichi pagani nordici, realizzate in forma di tre triangoli compenetranti per simboleggiare l'intervento del dio Odino!

\begin{figure}[!t]
\centering
\includegraphics[clip=,width=12cm]{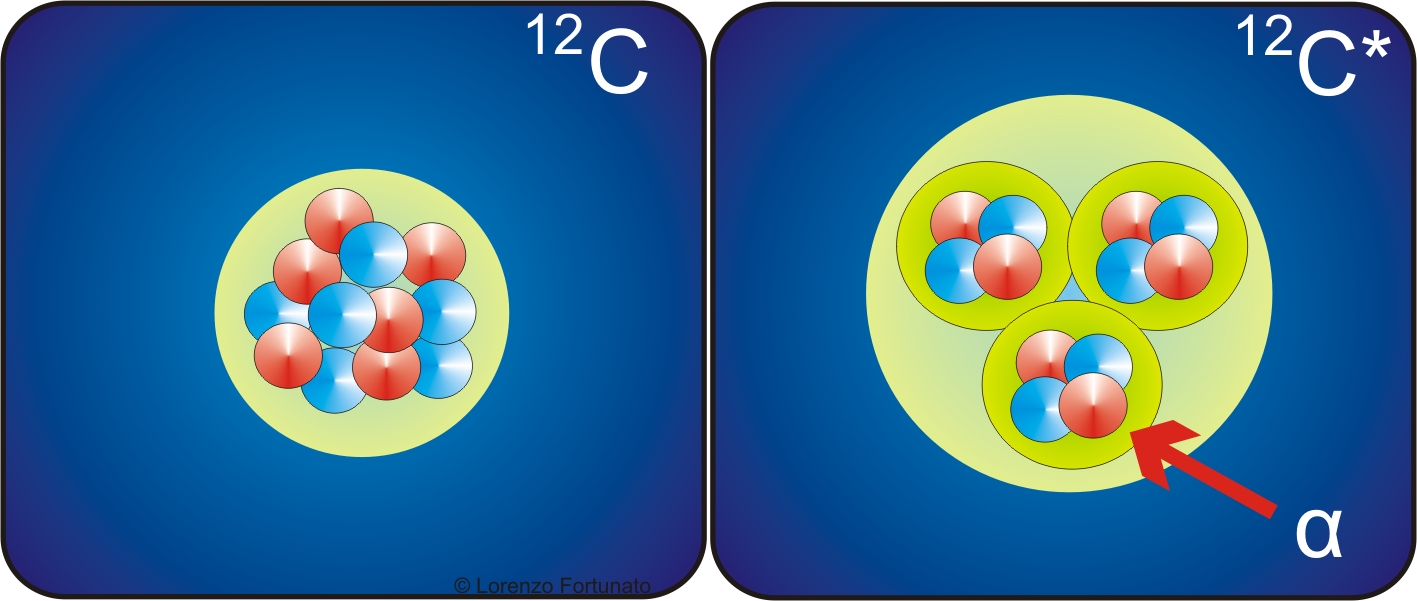}
\caption{Il nucleo di carbonio-12, l'isotopo pi\ù abbondante in natura di questo elemento necessario alla vita, \è formato da 6 protoni e 6 neutroni. Nel suo stato fondamentale (sinistra) questi sono quasi certamente raggruppati in un nocciolo unico, ma la struttura di alcuni stati eccitati (come il famoso stato di Hoyle a circa 7.5 MeV di energia di eccitazione), probabilmente corrisponde ad avere tre particelle a distinte e parzialmente sovrapposte (immagine a destra).}
\label{carb}
\end{figure}

Il fatto che questi nuclei siano poco legati gioca un ruolo fondamentale in varie reazioni nucleari. Ad energie che a noi sembrano altissime, ma che non sono troppo elevate per i sistemi nucleari, come quelle che si possono instaurare in certi regimi nelle stelle, questi nuclei hanno vita breve e tendono a spaccarsi lasciando scappare facilmente le particelle che formano il loro alone. Altri nuclei, possono presentare una struttura ``esotica'' non gi\à nel loro stato fondamentale, bens\ì in qualche stato eccitato: \è il caso del carbonio-12, l'elemento chiave della vita sul nostro pianeta! Il carbonio si forma nelle stelle: negli astri come il nostro Sole la stragrande maggioranza del combustibile \è idrogeno, ma vi \è una percentuale considerevole di elio-4 (particella $\alpha$). Quando due particelle $\alpha$ collidono non formano un nucleo legato: il berillio-8 (Z=4) non esiste in natura, anche se alle temperature del Sole pu\ò, per cos\ì dire, resistere per una frazione brevissima di secondo in uno stato detto risonante. Immaginate due particelle $\alpha$ come biglie di ferro cosparse di grasso vischioso (del tipo di quello che si usava una volta per gli ingranaggi delle biciclette, dopo essersi dimenticati il barattolo aperto per un paio di giorni) che si scontrano ad una velocit\à elevata, ma non troppo forte: le particelle sono di ``ferro'' in quanto alle energie in gioco non sono certo in grado di rompersi o deformarsi, mentre il grasso rappresenta la forza nucleare forte che \è a corto raggio. Ebbene nel momento della collisione possiamo immaginare che i due strati di grasso si tocchino e tengano unito il sistema delle due biglie per una piccola frazione di secondo (mentre se fossero state pulite, cio\è senza interazione, avrebbero rimbalzato senza attrarsi mai).
Ebbene, se nel tempo effimero del ``vischioso abbraccio'' una terza particella $\alpha$ passasse da quelle parti (fatto che avviene di rado, ma non tanto quanto si possa pensare date le alte densit\à dell'ambiente stellare) si potreebe formare uno stato legato del carbonio-12 (Z=6). E' questo uno stato molto particolare, detto stato di Hoyle, dal nome del fisico che ha proposto questa reazione: \è legato, ma possiede un'energia di eccitazione di 7.65MeV e molto probabilmente ha una struttura ``a cluster'', ovvero dobbiamo immaginarcelo non come un unico grappolo di dodici biglie, ma come un sistema di tre piccoli grappoli (nuclei di elio-4), di quattro biglie ciascuno. La figura \ref{carb} illustra, in maniera schematica, la struttura dello stato fondamentale e dello stato di Hoyle. Ora, \è possibile che lo stato di Hoyle decada riemettendo una particella $\alpha$, e il rimanente berillio-8, non essendo legato, si sfasci a sua volta in altre due particelle $\alpha$, ma \è anche possibile che decada per emissione di un raggio $\gamma$ che, sottraendo energia al sistema, lo converte nel suo stato fondamentale. Ecco cos\ì spiegata la produzione di carbonio (e di molti elementi pesanti) nelle stelle, che ha come stato di passaggio (ovvero di ``door-way state'' in inglese) proprio una struttura nucleare esotica. I nuclei con massa atomica esattamente divisibile per 4, ovvero $A=4N$, vengono detti $\alpha$-coniugati e possono presentare, nel loro stato fondamentale o in un sottoinsieme dei loro stati eccitati, il fenomeno della clusterizzazione-$\alpha$. In sequenza sono 4He, 8Be, 12C, 16O, 20Ne, 24Mg, 28Si, 32S, etc. e il modello ad $\alpha$-clusters, ideato da Hafstad e Teller, sebbene schematico, mostra come la loro energia di legame sia in proporzione diretta al numero di legami tra le particelle $\alpha$. L'interazione tra le particelle $\alpha$ o pi\ù in generale tra clusters di varia composizione, che in linea di principio si potrebbe derivare dall'azione combinata di svariate interazioni tra i singoli nucleoni che le costituiscono, ricordano qualitativamente le deboli interazioni di Van der Waals che sono in opera, per esempio, tra atomi neutri e tendono a formare legami piuttosto deboli. 
Ci sono altri nuclei formati da clusters, come \è stato suggerito fin dagli anni '80, in particolare il litio-6 ed il litio-7 (Z=3) sono entrambi descrivibili come sistemi legati di una particella $\alpha$ ed un nucleo di deuterio o di tritio, rispettivamente. Queste sono vere e proprie "molecole nucleari" binarie, tenute insieme dalla forza nucleare forte. Si pu\ò pensare che siano l'analogo nucleare delle molecole ioniche diatomiche, che essendo formate da ioni di segno opposto sono tenute assieme dalla forza elettrostatica. Ma il parallelo con la chimica deve essere inteso solo in senso qualitativo: mentre gli ioni hanno cariche opposte, qui i due clusters hanno cariche positive (possono contenere solo protoni) che tendono a separarli, ed \è l'attrazione nucleare (vedi figura \ref{pot}) che li tiene legati.
Con il nostro gruppo di ricerca presso l'Universit\à di Padova e la sede di Padova dell'I.N.F.N. (Istituto Nazionale di Fisica Nucleare) e in collaborazione con i Laboratori Nazionali di Legnaro, abbiamo applicato un semplice modello a due clusters alla descrizione delle propriet\à di struttura e di reazione degli isotopi del litio, ottenendo un ottimo accordo coi dati sperimentali per esempio per ci\ò che riguarda le reazioni di break-up, cio\è di separazione nei due clusters costituenti o per reazioni come la fotoemissione e la cattura radiativa che hanno rilevanza nello studio dei modelli stellari \cite{nos}.

\begin{figure}[!t]
\centering
\includegraphics[clip=,width=10cm]{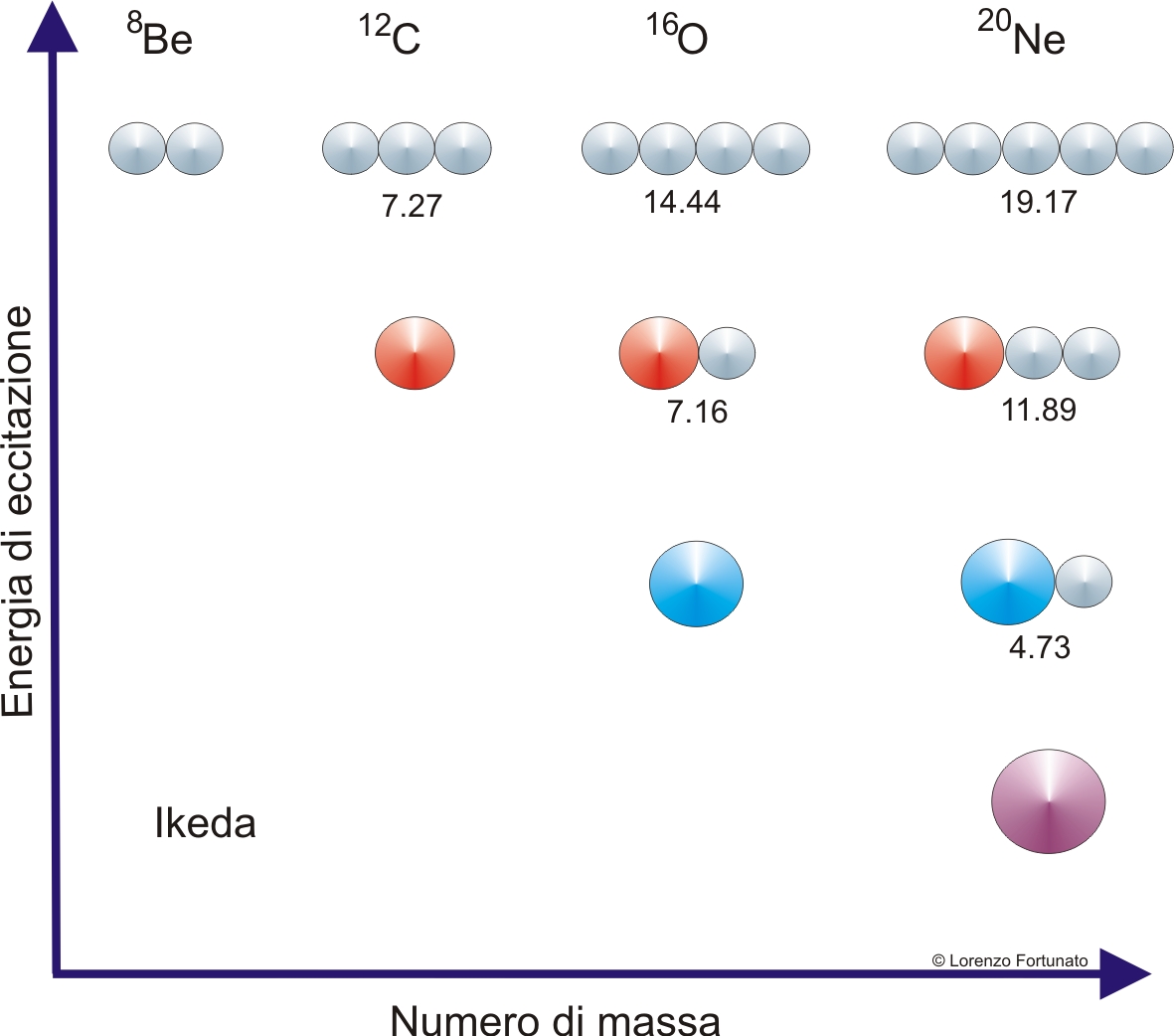}
\caption{Diagrammi di soglia di Ikeda per i primi quattro nuclei $\alpha$-coniugati: berillio-8, carbonio-12, ossigeno-16 e neon-20. I numeri indicati in corrispondenza dei disegni rappresentano le energie necessarie (in MeV) per decomporre un certo nucleo nei sottoinsiemi di clusters schematizzati con palline colorate: in grigio le particelle $\alpha$, in rosso il carbonio-12, in blu l'ossigeno-16 e in viola il neon-20. Ikeda sugger\ì nel 1968 che queste strutture si possano formare in vicinanza della soglia per emissione dei clusters, piuttosto che nello stato fondamentale, un fatto che \è stato pi\ù volte osservato sperimentalmente. Adattato da \cite{vonO}.}
\label{ikeda}
\end{figure}

La fisica dei clusters si \è evoluta molto negli ultimi anni: ad esempio si possono ipotizzare vere e proprie strutture molecolari \cite{vonO} in cui i clusters pi\ù compatti sono da intendersi come dei centri che vengono tenuti insieme dalla presenza di neutroni di valenza. Nelle molecole covalenti due atomi possono condividere un certo numero di elettroni per formare un orbitale molecolare (si pensi per esempio alla molecola di $H_2$ nella quale i due nuclei condividono i due elettroni che occupano un orbitale che li ingloba entrambi). Analogamente, tornando ai nuclei, due particelle $\alpha$ possono condividere un ulteriore neutrone per formare un legame nel berillio-9, creando una sorta di ``molecola nucleare covalente''. Teorie tipiche della fisica molecolare sono state adattate al caso nucleare: per esempio il metodo LCAO (Linear Combination of Atomic Orbitals) ha un suo analogo nel LCNO (Linear Combination of Nuclear Orbitals). 
\begin{figure}[!t]
\centering
\includegraphics[clip=,width=10cm]{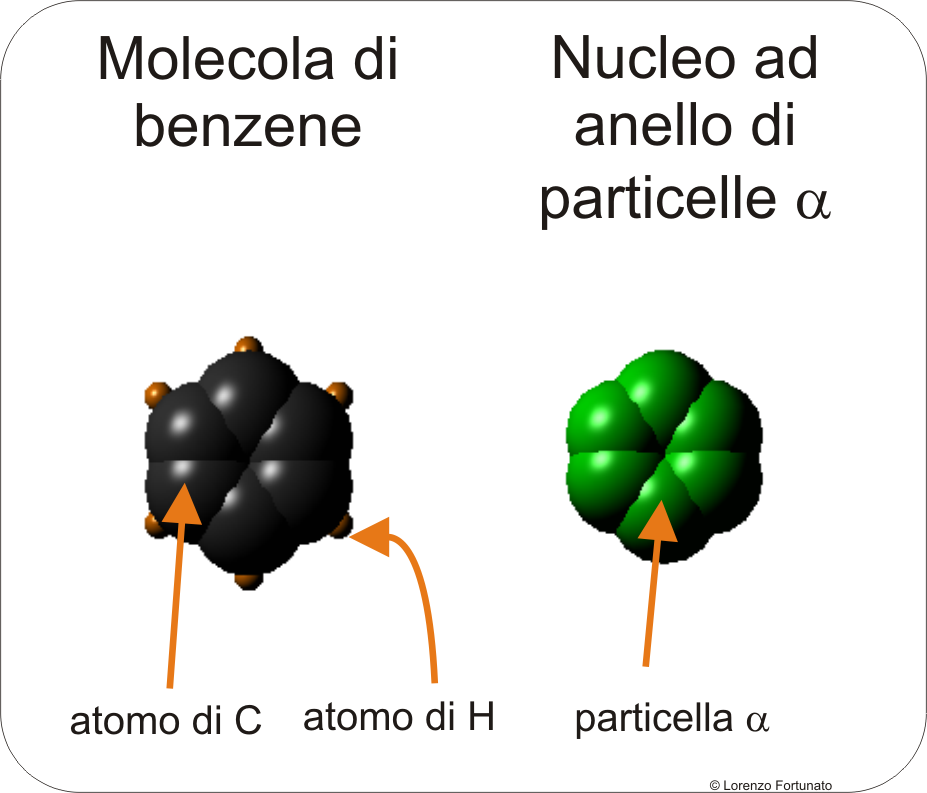}
\caption{La molecola di benzene (sinistra) \è composta da 6 atomi di carbonio che forniscono quattro elettroni ciascuno:  uno serve a tenere legato un atomo di idrogeno, e gli altri tre vengono messi in comune coi vicini. Dato che alcuni degli elettroni che legano l'anello non sono pi\ù attribuibili spazialmente ad un singolo atomo si parla di legame delocalizzato. Allo stesso modo si ipotizza che esistano nel nucleo di magnesio-24 con stati eccitati composti da sei particelle $\alpha$ (immagine a destra) e che negli isotopi più pesanti come magnesio-25, etc. possano esistere stati eccitati ad anello con neutroni delocalizzati. Per dovere di cronaca la dimensione di un tale sistema nucleare dovrebbe risultare circa centomila volte pi\ù piccola della molecola di benzene.}
\label{benz}
\end{figure}
Sono state avanzate ipotesi di strutture a 4, 5 e pi\ù clusters, con o senza neutroni di valenza e perfino anelli di particelle $\alpha$ disposte lungo un cerchio che, un po' come accade per il benzene (si veda la figura \ref{benz}), possono condividere un certo numero di neutroni delocalizzati! Molte di queste interessanti speculazioni, che sono raccolte in diagrammi detti di Ikeda (si veda la figura \ref{ikeda}) dal nome dello scienziato giapponese che propose che questi schemi molecolari semplificati sono da ricercarsi ad energie prossime a quelle di dissociazione nei clusters corrispondenti. Sebbene molte di queste ipotetiche strutture siano ancora prive di riscontro sperimentale, e probabilmente verranno smentite dalle misurazioni, molte altre sono state individuate e rimangono affascinanti contaminazioni tra chimica, fisica molecolare, geometria e fisica nucleare.
La clusterizzazione, cio\è la tendenza dei nucleoni a raggrupparsi in piccoli sottoinsiemi corrispondenti a nuclei molto stabili e compatti, esiste contemporaneamente al campo medio, ovvero alla tendenza dei nucleoni ad occupare una sorta di ``pozzo'' comune di energia uniforme e privo di sottostrutture. Questi due limiti, ``granulare'' uno ed ``uniforme'' l'altro, si manifestano di solito in diversi regimi di energia e di composizione in termini di protoni e neutroni. In particolare abbiamo discusso qui il caso dei nuclei leggeri, ma la formazione di $\alpha$-clusters \è importante anche all'altro estremo della carta dei nuclidi: nei nuclei instabili pesanti come l'uranio, il torio o il plutonio, che decadono emettendo una particella $\alpha$, si pu\ò ipotizzare che quest'ultima sia gi\à preformata all'interno del nucleo padre che tende a decadere. Lo studio dei clusters in nuclei pesanti e della loro emissione fornisce un altro aspetto intrigante della clusterizzazione, ma questo ci porterebbe lontano dalla nostra rotta originaria.

Un'altra propriet\à interessante dei nuclei instabili medio-pesanti ricchi di neutroni \è la formazione di una pelle di neutroni. Quando un certo isotopo possiede una forte sproporzione tra il numero di neutroni e il numero di protoni, le densit\à relative dei due costituenti tendono a calare in maniera assai differente nella zona superficiale: questo fa s\ì che si possa individuare una zona esterna in cui la densit\à dei protoni \è gi\à diminuita significativamente, ma la densit\à di neutroni \è ancora considerevole. Pertanto si forma una sorta di pelle, o di guscio, che contiene prevalentemente neutroni. La densit\à centrale di tutti i nuclei, dai pi\ù leggeri (esclusi quelli con massa molto piccola) ai pi\ù pesanti, \è praticamente costante e pari a circa 0.17 nucleoni/fm$^3$. Ci\ò \è dovuto alla saturazione del legame nucleare: essendo la buca attrattiva di figura \ref{pot} caratterizzata da corto raggio d'azione paragonabile alla grandezza delle particelle stesse, si ha che ogni nucleone forma legami solo con quelli a lui pi\ù prossimi e l'aggiunta di ulteriori nucleoni non influenza la densit\à del centro.
Il caso dell'alone e della pelle di neutroni sono differenti in quanto nel primo la densit\à delle particelle superficiali cala lentamente e su una grande regione di spazio, mentre nel secondo non c'\è quasi differenza nelle particolari forme in cui calano la densit\à protonica e neutronica alla superficie, ma i neutroni essendo pi\ù abbondanti continuano a godere della saturazione dei legami e mantengono la densit\à di saturazione a raggi pi\ù grandi, come illustrato in figura \ref{skin}.

Un fenomeno simile si poterebbe incontrare, in principio, anche dal lato della tavola dei nuclidi che \è ricco di protoni, ma in questo caso la repulsione elettrostatica tra i protoni, tutti carichi positivamente, impedisce la stabilit\à dei sistemi prima che si possa instaurare una vera e propria pelle di protoni.
La presenza di pelle di neutroni modifica il bilancio di alcune reazioni nucleari: ad esempio, in processi periferici, detti cos\ì perch\é i due nuclei in collisione si sfiorano appena, ci si aspetta che i processi di trasferimento di neutroni siano fortemente intensificati dalla presenza della pelle di neutroni, i quali non solo trovano energeticamente favorevole il passaggio da un nucleo ricco di neutroni ad un nucleo meno ricco, ma vengono anche favoriti nel loro trasferimento dall'alta densit\à superficiale e dalla vicinanza spaziale.

\section{CONCLUSIONE}

La ricchezza di manifestazioni che si nascondono in ogni nuovo sistema nucleare che viene studiato in laboratorio non si esaurisce con la carrellata di strani fenomeni descritti in queste pagine, ovvero aloni, clusterizzazione, pelle di neutroni, etc.. Nuove entusiasmanti scoperte sono all'orizzonte.

I fisici, come esploratori e cartografi, stanno esplorando ogni anfratto, ogni insenatura di queste terre esotiche che giacciono ai margini della stabilit\à dei sistemi nucleari, scoprendo, letteralmente di giorno in giorno, nuovi fenomeni, nuove manifestazioni della simmetria, nuovi effetti quantistici che rivoluzionano la nostra conoscenza del nucleo atomico. La fisica nucleare, e in particolare anche i fenomeni esotici di cui abbiamo discusso, \è strettamente interconnessa con tutte le altre discipline del sapere umano, e ha poco o nulla a che vedere con la guerra o l'inquinamento. Lo studio di queste bizzarrie del mondo nucleare ha allargato i nostri orizzonti, ci ha permesso di estendere e testare varie teorie sulla struttura del nucleo atomico, ma ha anche prodotto innumerevoli applicazioni ad altri campi, in maniera rimarchevole nell'astrofisica nucleare. Ha spinto la tecnologia di rivelazione di particelle cariche, neutroni e raggi $\gamma$ a livelli impensabili solo una ventina o trentina di anni fa e permette di sperare che le ricadute tecnologiche in altri campi, come ad esempio quello basilare delle terapie non invasive per i tumori basate sull'utilizzo di fasci accelerati di particelle, possano inglobare queste nuove conoscenze con vantaggi che possono riguardare la vita di tutti noi.

\section*{PROFILO DELL'AUTORE}
\begin{wrapfigure}{r}{50mm}
  \begin{center}
    \includegraphics[clip=,width=5cm]{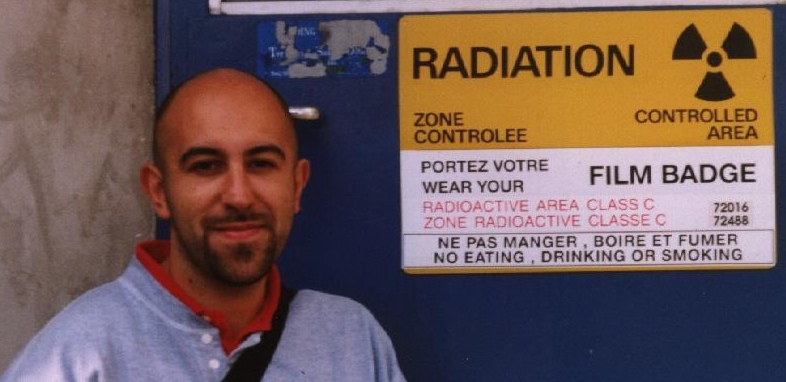}
  \end{center}
\end{wrapfigure}
Lorenzo Fortunato (Laurea Fisica presso l'Universit\à di Torino, 2000 e Dottorato di Ricerca in Fisica presso l'Universit\à Padova, 2003) dopo aver lavorato nelle universit\à di Padova, Siviglia (Spagna) e Gent (Belgio) \è ricercatore post-doc presso l'ECT* di Trento, un centro internazionale dedicato alla Fisica nucleare teorica. 
Le sue ricerche riguardano la struttura dei nuclei (modelli algebrici, modelli collettivi) e le reazioni nucleari (break-up, reazioni di trasferimento). Si interessa di fisica dei nuclei lontani dalla valle di stabilit\à, di eccitazione di modi collettivi, di soluzioni dell'hamiltoniana di Bohr, di nuclei a cluster, di sistemi a molti corpi in generale, dell'applicazione di algebre di Lie e teoria dei gruppi alla fisica nucleare, di fisica matematica e di molti altri argomenti come la sonoluminescenza e le energie rinnovabili. E' autore di oltre 50 articoli su riviste internazionali e relatore in altrettanti convegni e seminari.

\newpage
\begin{mdframed}[linewidth =2,roundcorner=30pt, backgroundcolor = Khaki, linecolor=MediumBlue]
\section*{I DECADIMENTI RADIOATTIVI}

\subsection*{Decadimento $\alpha$}
\begin{wrapfigure}{r}{60mm}
 \begin{center}
    \includegraphics[clip=,width=5cm]{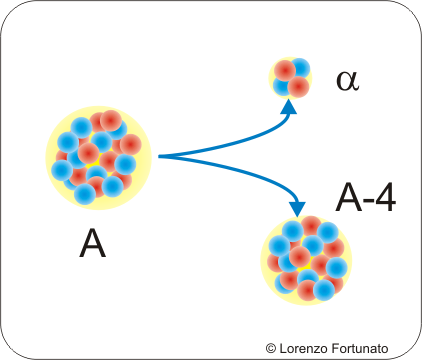}
  \end{center}
\end{wrapfigure}
I nuclei superpesanti ($Z>92$), molti nuclei pesanti ($Z\sim 90$ e $N >>Z$) ed alcuni nuclei pi\ù leggeri tendono a decadere emettendo una particella $\alpha$, ovvero un nucleo di elio-4 che contiene due protoni e due neutroni. Pertanto, nel decadimento da nucleo padre a nucleo figlio, si ha una riduzione di massa di quattro nucleoni $A\rightarrow (A-4) + \alpha$ e una maggiore stabilit\à del sistema finale. L'Americio-241 \è un isotopo radioattivo che decade con questa modalit\à e viene utilizzato in dosi minime nei rilevatori di fumo per allarmi antincendio. Decadendo, la particella $\alpha$ che viene emessa porta con s\é dell'energia cinetica che, nell'urto con alcune molecole d'aria le ionizza, determinando una piccola corrente elettrica tra due elettrodi. Il fumo altera questa corrente che viene rilevata da appositi circuiti elettrici. 
La radiazione $\alpha$ \è decisamente ionizzante, ma molto poco penetrante e viene in genere bloccata da pochi centimetri d'aria o alternativamente da strati molto sottili di materiali pi\ù densi come una tuta in polietilene dello spessore millimetrico o lo strato pi\ù esterno della nostra epidermide!
\subsection*{Decadimento $\beta$}
\begin{wrapfigure}{r}{60mm}
  \begin{center}
    \includegraphics[clip=,width=5cm]{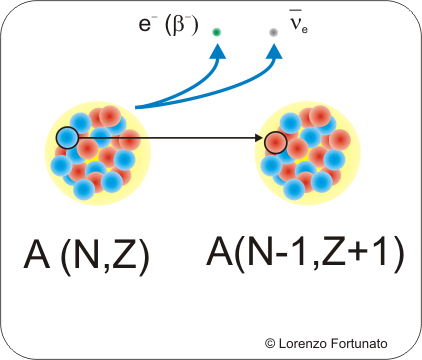}
  \end{center}
\end{wrapfigure}
Praticamente tutti i nuclei che si trovano al di fuori della valle di stabilit\à decadono con questa modalit\à che si presenta in due processi ``speculari'':  il decadimento $\beta-$ che consta nell'emissione di un elettrone ($e^-$) e di un antineutrino elettronico ($\bar \nu_e$) e il decadimento $\beta+$ che consta nell'emissione di un positrone ($e^+$ , l'antiparticella dell'elettrone con carica positiva) e di un neutrino elettronico ($\nu_e$). In questi decadimenti uno dei nucleoni del nucleo padre cambia natura trasformandosi da protone in neutrone nel primo caso o da neutrone in protone nel secondo. In sintesi
\begin{eqnarray}
A(N,Z) &\rightarrow & A(N+1,Z-1) + e^- +\bar \nu_e ~~ (dec. \beta-) \nonumber\\ 
A(N,Z) &\rightarrow & A(N-1,Z+1) + e^+ + \nu_e ~~ (dec. \beta+)\nonumber
\end{eqnarray}
da cui si pu\ò vedere che la carica elettrica totale si conserva nel processo di decadimento. Questo processo ``avvicina'' i nuclei instabili alla stabilit\à facendoli rotolare (in passi di un uno alla volta) dai pendii verso la vallata (si veda fig. \ref{valle}). Le radiazioni $\beta$ sono altamente ionizzanti e poco penetranti (ma molto più\ delle $\alpha$) e vengono in generate schermate con pochi centimetri di materiale solido di alto numero atomico (tipicamente piombo).
\subsection*{Decadimento $\gamma$}
\begin{wrapfigure}{r}{60mm}
  \begin{center}
   \includegraphics[clip=,width=5cm]{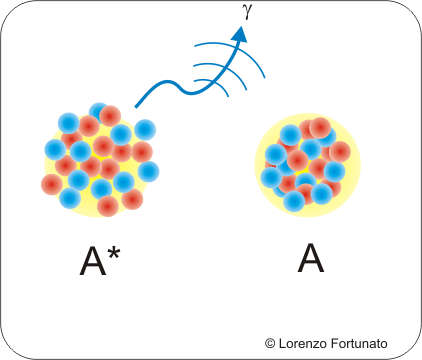}
  \end{center}
\end{wrapfigure}
Questo processo \è un riarrangiamento del nucleo verso una configurazione pi\ù stabile (cio\è ad energia pi\ù bassa) che non cambia la sua massa e non emette particelle massiccie, ma solo fotoni dovuti alla diseccitazione di uno stato quantico di energia elevata verso uno di energia inferiore, ovvero $A^*\rightarrow A + \gamma$. I fotoni $\gamma$ sono ``quanti''  o ``pacchetti'' di luce, ovvero onde elettromagnetiche come il calore, la luce visibile, le radioonde, le microonde, i raggi X, etc., con un'energia piuttosto alta (dell'ordine dei Mega elettron volt). Sono meno ionizzanti delle altre radiazioni, ma molto penetranti e difficili da schermare (alcuni metri di calcestruzzo o alcuni decimetri di piombo sono in genere sufficienti, ma dipende dall'energia).\medskip
\end{mdframed}

\end{document}